\documentclass[aps,prl,superscriptaddress,reprint]{revtex4-2}
\usepackage{siunitx}
\newcommand{\round}[2]{\num[round-mode=places,round-precision=#1]{#2}}
\usepackage{microtype}

\usepackage{xfp}

\usepackage{chngcntr}

\usepackage{latexsym}      
\usepackage{dcolumn}
\usepackage[hidelinks]{hyperref}
\hypersetup{
 colorlinks=false,
 citecolor=false,
 linkcolor=false,
 urlcolor=false,
 pdfpagemode=UseNone,
 pdfstartview=FitH}

\usepackage{scalerel,stackengine,amsmath}

\usepackage[eulergreek]{sansmath}
\usepackage{ upgreek }

\usepackage{filecontents}

\usepackage{booktabs}
\usepackage{dcolumn}
\newcolumntype{d}[1]{D{.}{.}{#1}}

\usepackage{bm}
\renewcommand{\vec}[1]{\text{\bfseries #1}}

\usepackage{xcolor}
\usepackage{dcolumn}

\usepackage{tikz}
\usetikzlibrary{math}
\usetikzlibrary{positioning}
\usetikzlibrary{shapes.geometric, arrows}
\usetikzlibrary{arrows.meta}
\usetikzlibrary{positioning}
\usetikzlibrary{shapes,arrows}
\usetikzlibrary{intersections}
\usetikzlibrary{svg.path}
\usepackage{pgfplots}
\pgfplotsset{
tick label style = {font=\sansmath\sffamily},
every axis label/.append style = {font=\sansmath\sffamily}
}

\usetikzlibrary{arrows}
\pgfplotsset{compat=newest}
\usepgfplotslibrary{fillbetween}
\usetikzlibrary{calc}
\def\centerarc[#1](#2)(#3:#4:#5)
{ \draw[#1] ($(#2)+({#5*cos(#3)},{#5*sin(#3)})$) arc (#3:#4:#5); }

\pgfplotscreateplotcyclelist{usualsuspectinner}{
{mark=*,c1},
{mark=triangle*,c4},
{mark=diamond*,gold},
{mark=halfcircle*,c11},
{mark=square*,c21}, 
{mark=halfsquare left*,c23},
{mark=pentagon*,c28},
      }
\pgfplotscreateplotcyclelist{usualsuspectouter}{
{mark=o,c1},
{mark=triangle,c4},
{mark=diamond,gold},
{mark=halfcircle,c11},
{mark=square,c21}, 
{mark=halfsquare left,c23},
{mark=pentagon,c28},
      }

\pgfplotsset{
    every first x axis line/.style={},
    every first y axis line/.style={},
    every first z axis line/.style={},
    every second x axis line/.style={},
    every second y axis line/.style={},
    every second z axis line/.style={},
    first x axis line style/.style={/pgfplots/every first x axis line/.append style={#1}},
    first y axis line style/.style={/pgfplots/every first y axis line/.append style={#1}},
    first z axis line style/.style={/pgfplots/every first z axis line/.append style={#1}},
    second x axis line style/.style={/pgfplots/every second x axis line/.append style={#1}},
    second y axis line style/.style={/pgfplots/every second y axis line/.append style={#1}},
    second z axis line style/.style={/pgfplots/every second z axis line/.append style={#1}}
}

\makeatletter
\def\pgfplots@drawaxis@outerlines@separate@onorientedsurf#1#2{%
    \if2\csname pgfplots@#1axislinesnum\endcsname
    \else
    \scope[/pgfplots/every outer #1 axis line,
        #1discont,decoration={pre length=\csname #1disstart\endcsname, post length=\csname #1disend\endcsname}]
        \pgfplots@ifaxisline@B@onorientedsurf@should@be@drawn{0}{%
            \draw [/pgfplots/every first #1 axis line] decorate {
                \pgfextra
                \pgfplotspointonorientedsurfaceabsetupfor{#2}{#1}{\pgfplotspointonorientedsurfaceN}%
                \pgfplots@drawgridlines@onorientedsurf@fromto{\csname pgfplots@#2min\endcsname}%
                \endpgfextra 
                };
        }{}%
        \pgfplots@ifaxisline@B@onorientedsurf@should@be@drawn{1}{%
            \draw [/pgfplots/every second #1 axis line] decorate {
                \pgfextra
                \pgfplotspointonorientedsurfaceabsetupfor{#2}{#1}{\pgfplotspointonorientedsurfaceN}%
                \pgfplots@drawgridlines@onorientedsurf@fromto{\csname pgfplots@#2max\endcsname}%
                \endpgfextra 
                };
        }{}%
    \endscope
    \fi
}%
\makeatother

\usepackage{pgfplotstable}

\pgfplotsset{select coords between index/.style 2 args={
    x filter/.code={
        \ifnum\coordindex<#1\fi
        \ifnum\coordindex>#2\fi
    }
}}

\pgfdeclareplotmark{rottriangle}{%
  \pgfpathmoveto{\pgfqpoint{0pt}{-\pgfplotmarksize}}%
  \pgfpathlineto{\pgfqpointpolar{30}{\pgfplotmarksize}}%
  \pgfpathlineto{\pgfqpointpolar{150}{\pgfplotmarksize}}%
  \pgfpathclose
  \pgfusepathqstroke
}%
\pgfdeclareplotmark{rottriangle*}{%
  \pgfpathmoveto{\pgfqpoint{0pt}{-\pgfplotmarksize}}%
  \pgfpathlineto{\pgfqpointpolar{30}{\pgfplotmarksize}}%
  \pgfpathlineto{\pgfqpointpolar{150}{\pgfplotmarksize}}%
  \pgfpathclose
  \pgfusepathqfillstroke
}

\tikzset{every picture/.style={font issue=\footnotesize},
         font issue/.style={execute at begin picture={#1\selectfont}}
        }

\usepackage{csvsimple}
\usepackage{changepage}
\usepackage[tbtags]{mathtools}
\usepackage{siunitx}[=v2]
\usepackage{csquotes}
\usepackage{enumerate}
\usepackage{enumitem}
\usepackage{filecontents}

\usepackage[
	nonumberlist, 
	acronym, 
	nomain,
	nopostdot,
]{glossaries}

\usepackage{color, colortbl}

\usepackage[utf8]{inputenc}
\usepackage{csvsimple}
\usepackage{tikz}
\usepackage{nicefrac}

\usepackage{changepage}
\usepackage{color, colortbl}

\usepackage{cleveref}

\usepackage{blindtext}
\usepackage{amssymb}

\let\vec\mathbf
\usepackage{nicefrac}

\usepackage{multirow}
\usepackage{tabularx}
\usepackage{array}
\newcolumntype{L}[1]{>{\raggedright\let\newline\\\arraybackslash\hspace{0pt}}m{#1}}
\newcolumntype{C}[1]{>{\centering\let\newline\\\arraybackslash\hspace{0pt}}m{#1}}
\newcolumntype{R}[1]{>{\raggedleft\let\newline\\\arraybackslash\hspace{0pt}}m{#1}}

\usepackage{filecontents}

\usepackage[normalem]{ulem}

\newcommand\red[1]{\textcolor{black}{#1}}

\usepackage{amssymb}

\DeclareMathOperator{\Rey}{Re}
\DeclareMathOperator{\Pe}{Pe}
\DeclareMathOperator{\St}{St}
\DeclareMathOperator{\Ta}{Ta}
\DeclareRobustCommand{\orderof}{\ensuremath{\mathcal{O}}}

\newacronym{md}{MD}{molecular dynamics}
\newacronym{dc}{DC}{direct-current}
\newacronym[plural=PFCs,firstplural=parabolic flight campaigns (PFCs)]{pfc}{PFC}{Parabolic Flight Campaign}
\newacronym{fft}{FFT}{Fast Fourrier Transform}
\newacronym{cad}{CAD}{Computer Assisted Design}
\newacronym{ptfe}{PTFE}{polytetrafluoroethylene}
\newacronym{ps}{PS}{polystyrene}
\newacronym{esa}{ESA}{European Space Agency}
\newacronym{dlr}{DLR}{German Aerospace Center, abbreviated from German \textit{Deutsches Zentrum f\"{u} Luft- und Raumfahrt e.V.}}
\newacronym{liggghts}{LIGGGHTS}{\acrshort{lammps} Improved for General Granular and Granular Heat Transfer Simulations}
\newacronym{rpm}{rpm}{revolutions per minute}
\newacronym{rise}{RISE}{Research Internships in Science and Engineering}
\newacronym{daad}{DAAD}{German Academic Exchange Service, abbreviated from German \textit{Deutscher Akademischer Austauschdienst}}

\newacronym{mcr}{MCR}{modular compact rheometer}
\newacronym{mfc}{MFC}{Mass Flow Controller}

\newacronym{mct}{MCT}{mode-coupling theory}
\newacronym{itt}{ITT}{integration through transients}
\newacronym{gitt}{GITT}{granular integration through transient} 
\newacronym{kd}{K-D}{Krieger-Dougherty}

\newacronym{rcp}{rcp}{random close packing}
\newacronym{rlp}{rlp}{random loose packing}
\newacronym{si}{SI}{Supplementary Information}

\definecolor{c1}{rgb}{0.7068574918274737, 0.11027871818526241, 0.2747061222663145}%
\definecolor{c2}{rgb}{0.6780437401750237, 0.1857033496010309, 0.10475664313058389}%
\definecolor{c3}{rgb}{0.5819819292481591, 0.28135510723834917, 0.10365807489974799}%
\definecolor{c4}{rgb}{0.5234892407222805, 0.3183040932830017, 0.10308831855626377}%
\definecolor{c5}{rgb}{0.4801916866157751, 0.3399605081271155, 0.10271364194308724}%
\definecolor{c6}{rgb}{0.44359832760663015, 0.3553988979685888, 0.10242748858902176}%
\definecolor{c7}{rgb}{0.40913672397205286, 0.36794412723413256, 0.10218299803254591}%
\definecolor{c8}{rgb}{0.37320211354661986, 0.37925108836394167, 0.10195327596798023}%
\definecolor{c9}{rgb}{0.33135856975649947, 0.3904277497779026, 0.10171733001337946}%
\definecolor{c10}{rgb}{0.2751398809612443, 0.40252912490464393, 0.10145175401497963}%
\definecolor{c11}{rgb}{0.17705545280426335, 0.4169982861326329, 0.10112016988851782}%
\definecolor{c12}{rgb}{0.10370113411519366, 0.41987331701008007, 0.20784080256789766}%
\definecolor{c13}{rgb}{0.10672620229256541, 0.415683426104876, 0.2824387977867041}%
\definecolor{c14}{rgb}{0.10896052083074859, 0.412482459879469, 0.3262929409150867}%
\definecolor{c15}{rgb}{0.11082049183583692, 0.4097465781098524, 0.3585658834499722}%
\definecolor{c16}{rgb}{0.1125289398766243, 0.40717495121262287, 0.38576967807315987}%
\definecolor{c17}{rgb}{0.11424591411818866, 0.4045324469356048, 0.41127417284765605}%
\definecolor{c18}{rgb}{0.11613475317261168, 0.4015563987219263, 0.4376207984793854}%
\definecolor{c19}{rgb}{0.11843118069754927, 0.3978377246048391, 0.46769849466738594}%
\definecolor{c20}{rgb}{0.1215889511379443, 0.3925371130865723, 0.5062751486798138}%
\definecolor{c21}{rgb}{0.12675667510032929, 0.38336665224260497, 0.564172135389151}%
\definecolor{c22}{rgb}{0.13823697094516182, 0.36050804998003744, 0.6775165660716256}%
\definecolor{c23}{rgb}{0.2929229731919379, 0.2517707788576924, 0.9106622939003164}%
\definecolor{c24}{rgb}{0.5044517963791875, 0.15823641315504755, 0.8374462940274711}%
\definecolor{c25}{rgb}{0.5743505600217842, 0.14563681653078617, 0.7277475942695497}%
\definecolor{c26}{rgb}{0.6114384687170589, 0.13753415571759905, 0.6502354669913063}%
\definecolor{c27}{rgb}{0.6363457404298491, 0.13141740987926892, 0.586320806947322}%
\definecolor{c28}{rgb}{0.6557132733878439, 0.12622710616005786, 0.5268534496956088}%
\definecolor{c29}{rgb}{0.6725336221941806, 0.12137095721249477, 0.4649001891758376}%
\definecolor{c30}{rgb}{0.688593421429631, 0.11639483989072555, 0.39157406584043325}%
\definecolor{brickred}{rgb}{0.8, 0.25, 0.33}%
\definecolor{darkorange}{rgb}{1.0, 0.55, 0.0}%
\definecolor{persiangreen}{rgb}{0.0, 0.65, 0.58}%
\definecolor{persianindigo}{rgb}{0.2, 0.07, 0.48}%
\definecolor{cadet}{rgb}{0.33, 0.41, 0.47}%
\definecolor{turquoisegreen}{rgb}{0.63, 0.84, 0.71}%
\definecolor{sandybrown}{rgb}{0.96, 0.64, 0.38}%
\definecolor{blueblue}{rgb}{0.0, 0.2, 0.6}%
\definecolor{ballblue}{rgb}{0.13, 0.67, 0.8}%
\definecolor{greengreen}{rgb}{0.0, 0.5, 0.0}%
\definecolor{forestgreen}{rgb}{0.0, 0.42, 0.24}
\definecolor{amber}{rgb}{1.0, 0.75, 0.0}%
\definecolor{goldenrod}{rgb}{0.85, 0.65, 0.13}%
\definecolor{gold}{rgb}{0.86, 0.71, 0.23}%
\definecolor{tiffanyblue}{rgb}{0.04, 0.73, 0.71}%
\definecolor{bittersweet}{rgb}{1.0, 0.44, 0.37}%

\definecolor{purpleheart}{rgb}{0.41, 0.21, 0.61}%
\definecolor{richblack}{rgb}{0.0, 0.25, 0.25}
\definecolor{royalazure}{rgb}{0.0, 0.22, 0.66}
\definecolor{sacramentostategreen}{rgb}{0.0, 0.34, 0.25}

\definecolor{tyrianpurple}{rgb}{0.4, 0.01, 0.24}
\definecolor{redviolet}{rgb}{0.78, 0.08, 0.52}
\definecolor{vividburgundy}{rgb}{0.62, 0.11, 0.21}
\definecolor{amaranth}{rgb}{0.9, 0.17, 0.31}

\definecolor{ca}{RGB}{127,59,8}%
\definecolor{cb}{RGB}{179,88,6}%
\definecolor{cc}{RGB}{224,130,20}%
\definecolor{cd}{RGB}{253,184,99}%
\definecolor{ce}{RGB}{254,224,182}
\definecolor{cf}{RGB}{216,218,235}%
\definecolor{cg}{RGB}{178,171,210}%
\definecolor{ch}{RGB}{128,115,172}%
\definecolor{ci}{RGB}{84,39,136}%
\definecolor{cj}{RGB}{45,0,75}%

\def\Qtouconvwithcylinder{0.010306294}

\def\Di{0.024}
\def\Do{0.050}
\def\lengthL{0.038}

\pgfmathsetmacro{\deltaval}{\Do/\Di}
\pgfmathsetmacro{\kbval}{\deltaval/(\deltaval-1)}
\pgfmathsetmacro{\knval}{(2*\deltaval*\deltaval)/(\deltaval*\deltaval-1)}

\def\PhiQtwo{0.598567996}
\def\PhiQtwopointthree{0.592906265}
\def\PhiQtwopointfive{0.589554029}
\def\PhiQthree{0.582289939}
\def\PhiQthreepointfive{0.576218112}
\def\PhiQfive{0.562410696}

\def\particlediameter{0.000175}
\def\particlemass{7.0154e-9}

\begin{document}
\title{\textbf{Rheological regimes in agitated granular media under shear}}

\author{Olfa D'Angelo}
\affiliation{Institut Sup\'{e}rieur de l'A\'{e}ronautique et de l'Espace (ISAE-SUPAERO), Universit\'{e} de Toulouse, Toulouse, France}
\affiliation{Institute for Multiscale Simulation, Erlangen-N\"{u}rnberg Universit\"{a}t, Cauerstra\ss{}e 3, 91058 Erlangen, Germany}
\affiliation{Institute of Materials Physics in Space, German Aerospace Center (DLR), Linder H\"{o}he, 51170 Cologne, Germany}
 
\author{Matthias Sperl}
\affiliation{Institute of Materials Physics in Space, German Aerospace Center (DLR), Linder H\"{o}he, 51170 Cologne, Germany}
\affiliation{Institute for Theoretical Physics, University of Cologne,  Z\"{u}lpicher Stra\ss{}e 77, 50937 Cologne, Germany}

\author{W. Till Kranz}
\affiliation{Institute of Materials Physics in Space, German Aerospace Center (DLR), Linder H\"{o}he, 51170 Cologne, Germany}
\affiliation{Institute for Theoretical Physics, University of Cologne,  Z\"{u}lpicher Stra\ss{}e 77, 50937 Cologne, Germany}

\begin{abstract}

Agitated granular media have a rich rheology:
they exhibit Newtonian behavior at low shear rate and density,
develop a yield stress at high density,
and cross over to Bagnoldian shear thickening when 
sheared rapidly---making them challenging to encompass in one theoretical framework.
We measure the rheology of air-fluidized glass particles,
spanning five orders of magnitude in shear rate.
By comparing fluidization-induced to Brownian agitation,
we show that
all rheological regimes can be delineated by two dimensionless numbers---the Peclet number, $\Pe$,
and the ratio of shear-to-fluidization power, $\Pi$---and propose a constitutive relation that captures
all flow behaviors, qualitatively and quantitatively,
in one unified framework.
\end{abstract}

\maketitle

Systems out of equilibrium, 
which evolve or remain in a steady state through energy exchange with their environment, 
are widespread, yet poorly understood from a fundamental point of view. 
 Such systems can be found at many length-scales, from biological matter, to colloidal suspensions, emulsions, 
 up to traffic flow or even star-forming regions.
 In granular media,
constant agitation counteracts the particles' dissipative nature, transforming static
granular solids into dynamic granular fluids with a non-zero granular
temperature $T$, i.e., a finite mean kinetic energy per particle~\cite{Geldart1973, Abate2006}. 
 From industrial reactors~\cite{Werther2007,Guo2023} to geophysical flows~\cite{Wilson1984, Melosh1996, Roche2004, Breard2023},
 granular media under constant agitation are widespread,
but continue to be challenging to understand and describe theoretically.

Measuring the rheology of agitated granular media 
has yielded seemingly contradictory results. 
Considering a wide range of geometries
and agitation mechanisms (air-fluidized~\cite{Roche2004, Kottlan2018, Brzinski2010, Breard2023},
suspended in a liquid flow~\cite{Guazzelli2018,Duru2002,Gibilaro2007,Ancey1999}, 
tapped or vibrated~\cite{Dijksman2011,Hanotin2012,Zhang2018,Gnoli2016}, acoustically fluidized~\cite{Melosh1996,VanDerElst2012,Conrad2013}, 
combined air flow and vibrations~\cite{Guo2023}),
some studies report Newtonian rheology~\cite{Hagyard1966,Koos2012,Francia2021} 
while others find shear thinning~\cite{Bouwhuis1961,Tardos1998,Chen2021}, 
and yet others shear thickening (Bagnoldian) 
rheology~\cite{Bagnold1954, Brzinski2010, Gnoli2016}---or crossovers between these behaviors \cite{Schugerl1961,Hobbel1985,Gottschalk1986,Dijksman2011,Kottlan2018,Young2021,VanDerElst2012,Hanotin2012,Ancey1999}. 
Yet, a succinct and comprehensive constitutive equation for agitated granular fluids, describing the stress tensor, $\Sigma$, as a function of shear rate, $\dot\gamma$, remains elusive.

In this Letter, we demonstrate that the rheology of agitated granular
media actually encompasses all three regimes: Newtonian, shear
thinning, and Bagnoldian shear thickening. 
By comparing fluidization-induced agitation to Brownian agitation,
we extend arguments pertaining to colloidal suspensions 
to physically explain each regime.
We show that the regime transitions are controlled by two
dimensionless numbers: the Peclet number, $\Pe$,
and the ratio of shear-to-fluidization power, $\Pi$.
Finally, we show that the \gls{gitt}
\cite{Kranz2018, Kranz2020}---developed from \gls{mct}, 
a first-principles-based theory 
for glassy dynamics \cite{Fuchs2002, Janssen2018}---captures 
in one theoretical framework the complex rheology of agitated granular media.

\begin{figure}[tb]
\centering
\begin{tikzpicture}
\sisetup{detect-all}
\normalfont\sffamily
\sansmath\sffamily

\begin{loglogaxis}[
          xlabel = {$\Omega$~/\si{\per\second}},
          ylabel={$\sigma$~/\si{\pascal}},
          width=0.98\linewidth, height=0.65\linewidth,
          xmin=3e-4,
          xmax=3e2,
          ymin=5e-3,
          ymax=3e2,
          mark size=1.75pt,
          mark options={solid},
	area legend,
	legend columns=3,
    legend style={
			only marks,
			font=\footnotesize\sffamily,
			fill=none,
		    legend cell align=left,
		    at={(0.01,0.985)},
		    anchor=north west,
		     /tikz/column 2/.style={column sep=6pt},
		     /tikz/column 4/.style={column sep=6pt},
		     /tikz/column 6/.style={column sep=6pt},
		     /tikz/column 8/.style={column sep=6pt},
          	}
    ]
 
\foreach \PF/\marque/\col/\variable/\spec in {
\PhiQtwo/*/c25/2_2/,
\PhiQtwopointthree/triangle*/c22/2.3_1/,
\PhiQtwopointfive/diamond*/c10!90!yellow/2.5_7/,
\PhiQthree/pentagon*/gold!80!c4/3_1/only marks,
\PhiQthreepointfive/square*/c4/3.5_1/,
\PhiQfive/rottriangle*/c1/5_1/}{
    \edef\theoryplottemp{\noexpand\addplot [
    color=\col,
    mark=\marque,
\spec,
    line legend,
    error bars/.cd, 
    	y dir=both, 
    	y explicit,
		error bar style={color=\col, solid}
    ]
    table[
    	col sep=comma,
        x expr = (\noexpand\thisrow{n_up}*2*pi)/60, 
        y expr = \noexpand\thisrow{M_up}*1e-6/(2*pi*\lengthL*(\Di/2)^2),
        y error plus expr=\noexpand\thisrow{stdM_up}*1e-6/(2*pi*\lengthL*(\Di/2)^2),
        y error minus expr=\noexpand\thisrow{stdM_up}*1e-6/(2*pi*\lengthL*(\Di/2)^2),
    ]
	{Olfa_data/Q\variable.csv};
	\noexpand\addlegendentry{\round{3}{\PF} }
	}\theoryplottemp
}

\foreach \marque/\col/\variable/\spec in {
o/c25/2_2/only marks, 
triangle/c22/2.3_1/only marks,
diamond/c10!90!yellow/2.5_7/only marks, 
pentagon/gold!80!c4/3_1/only marks,
square/c4/3.5_1/only marks,
rottriangle/c1/5_1/only marks}{
    \edef\theoryplottemp{\noexpand\addplot [
    forget plot,
    color=\col!80!black,
    mark=\marque,
	\spec, thick,
    error bars/.cd, 
    	y dir=both, 
    	y explicit,
		error bar style={color=\col!80!black, solid}
    ]
    table[
        col sep=comma,
        x expr = \noexpand\thisrow{n_down}*1*((2*pi)/60), 
        y expr = \noexpand\thisrow{M_down}*1e-6/(2*pi*\lengthL*(\Di/2)^2),
        y error plus expr=\noexpand\thisrow{stdM_down}*1e-6/(2*pi*\lengthL*(\Di/2)^2),
        y error minus expr=\noexpand\thisrow{stdM_down}*1e-6/(2*pi*\lengthL*(\Di/2)^2),
    ]
	{Olfa_data/Q\variable.csv};
	}\theoryplottemp
}
\foreach \marque/\col/\variable in {
pentagon/gold!80!c4/3_1}{
    \edef\theoryplottemp{\noexpand\addplot [
    forget plot,
    color=\col,
	no marks,
]
    table[
        col sep=comma,
        x expr = (\noexpand\thisrow{n_down}*2*pi)/60, 
        y expr = \noexpand\thisrow{M_down}*1e-6/(2*pi*\lengthL*(\Di/2)^2),
        y error plus expr=\noexpand\thisrow{stdM_down}*1e-6/(2*pi*\lengthL*(\Di/2)^2),
        y error minus expr=\noexpand\thisrow{stdM_down}*1e-6/(2*pi*\lengthL*(\Di/2)^2),
    ]
	{Olfa_data/Q\variable.csv};
	}\theoryplottemp
}

\begin{scope}[xshift=-8pt, yshift=10pt]
\draw [gray] (axis cs:3e2,3e0)
-- (axis cs:8e1,3e0) node [midway, below]{$\propto\Omega^2$}
-- (axis cs:3e2,3e1) 
-- cycle;
\end{scope}
\begin{scope}[xshift=1pt, yshift=16pt]
\draw [gray] (axis cs:3e-2,2e-3)
-- (axis cs:3e-1,2e-3) node [midway, above, xshift=4pt]{$\propto\Omega$}
-- (axis cs:3e-1,2e-2) 
-- cycle;
\end{scope}

\coordinate (insetPosition) at (rel axis cs:1,0.01);
\end{loglogaxis}

\begin{axis}[font=\footnotesize,
at = (insetPosition), anchor=outer south east, 
          xlabel = {$u$~/\si{\m\per\s}},
          ylabel={$\varphi$},
	x label style={at={(rel axis cs:0.56,-0.05)},anchor=south},
    y label style={at={(rel axis cs:0,1)},rotate=-90,anchor=north west},
          width=0.467\linewidth, height=0.35\linewidth,
          xmax=0.065,
          ymin=0.535,
	fill=white,
	scaled x ticks = false,
	x tick label style={
		/pgf/number format/precision=4
		},
          mark size=1.75pt,
          mark options={solid},
	area legend,
	legend columns=3,
    legend style={
		    legend cell align=left,
		    at={(0.5,-0.21)},
		    anchor=north,
		     /tikz/column 2/.style={column sep=14pt}
          	}
    ]

\addplot
[  		
	forget plot,
    lightgray,
	thick,
    no marks, 
    domain=0.015:0.06,
]
{ (91960*x)^(-0.068) }; 

\addplot    [
	c25, mark = *,
	only marks,
	]
coordinates{
(2*\Qtouconvwithcylinder, \PhiQtwo)
        };
        
\addplot    [
	c22, mark = triangle*,
	only marks,
	]
coordinates{
(2.3*\Qtouconvwithcylinder, \PhiQtwopointthree)
        };
     
\addplot    [
	c10!90!yellow, mark = diamond*,
	only marks,
	]
coordinates{
(2.5*\Qtouconvwithcylinder, \PhiQtwopointfive) 
        };
        
\addplot    [
	gold!80!c4, mark = pentagon*,
	only marks,
	]
coordinates{
(3*\Qtouconvwithcylinder, \PhiQthree)
        };
        
\addplot    [
	c4, mark = square*,
	only marks,
	]
coordinates{
(3.5*\Qtouconvwithcylinder, \PhiQthreepointfive)
        };

\addplot    [
	c1, mark = rottriangle*,
	only marks,
	]
coordinates{
(5*\Qtouconvwithcylinder, \PhiQfive)
        };

\end{axis}

\end{tikzpicture}
\caption{\label{fig:experimental}
Steady state flow curves, $\sigma (\varOmega)$, for air-fluidized glass
beads: shear stress, $\sigma$, \textit{vs.}\ angular velocity, $\varOmega$, of the inner cylinder
in a Taylor-Couette shear cell. Each flow curve corresponds to 
a packing fraction, $\varphi$, given in the legend.
Filled marks correspond to the upward sweep, empty marks to the downward sweep in $\varOmega$.
Lines are a guide to the eye. Inset: Global packing fraction
\textit{vs.}\ fluidization air flow velocity, $\varphi (u)$; the gray line
is a fit to experimental data (details in Appendix~I).  }
\end{figure}
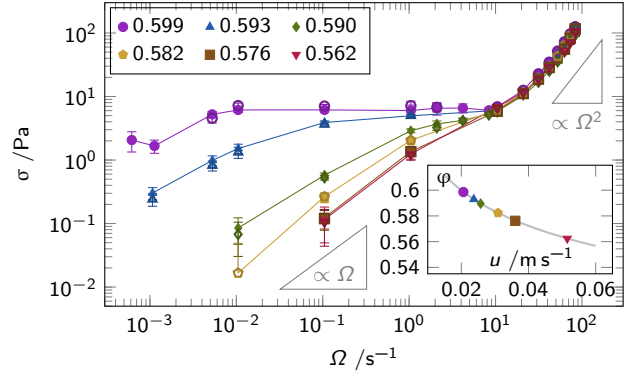

\textit{Experiments.---}We measure the steady-state rheology of
air-fluidized glass particles 
(glass bulk density
$\rho_p = \SI{2.5}{\g\per\cm\cubed}$, diameter $d \in $
[150-200]\,\si{\micro\m}, sample mass $M = \SI{170}{\g}$, Geldart group
B~\cite{Geldart1973}), and vary the fluidization gas flow velocity,
$u$. The single fluidization control parameter is $u$: the bed expands
with increasing $u$, which lowers the global packing fraction,
$\varphi$, (see Fig.\,\ref{fig:experimental} inset) and, at the same time,
increases agitation.  While $\varphi(u)$ characterizes the fluidized
bed's density, its agitation can be characterized by the injected
power density, $\Pi_\mathrm{f}(u) = \rho_p\varphi ug$ ($g$
gravitational acceleration).
In the unsheared bed, Stokes numbers 
$\St \gg 1$ \footnote{$\St := \rho_p d u / \eta_f $,
with the air's viscosity, $\eta_f$, and air flow velocity, $u$, are $\orderof (10^3)$.}, such
that particles are not over-damped by the surrounding fluid (air), but
driven by their inertia, forming a non-Brownian dry suspension.

The rheology of our air-fluidized granular bed is measured in a wide
gap Taylor-Couette shear cell (Anton Paar \acrshort{mcr}-102, coaxial
cylinders, geometry details in Appendix~I). The torque, $\mathcal M$, for
a fixed inner cylinder's angular velocity, $\varOmega$, is recorded
and converted to stress, $\sigma = \mathcal M/ 2\pi L R_i^2$ ($L$ and
$R_i$ inner cylinder height and radius, respectively).  Steady state
follows long transients (on the order of hours for slow shear).
Flow curves, $\sigma (\varOmega | \varphi )$, are plotted in
Fig.\,\ref{fig:experimental}, spanning five orders of magnitude in
angular velocity for a range of packing densities.

\textit{Rheological regimes.---}In Figure~\ref{fig:experimental}, we see
that an air-fluidized granular bed may behave as a Newtonian fluid at
low $\varphi$ and $\varOmega$, as a shear thickening fluid at high
$\varOmega$, or as a shear thinning fluid at intermediate $\varOmega$
and at high $\varphi$.  Similar regimes have been observed in
molecular glasses and colloids~\cite{Brader2009,Stickel2005,
  Wagner2021book}.
We will argue below that while Bagnoldian shear thickening is
genuinely granular, the Newtonian and shear thinning behaviors,
including evidence of an incipient granular glass transition, may be
understood in terms that generalize concepts developed for equilibrium
fluids, highlighting a universality of fluid-like, amorphous materials
across scales.

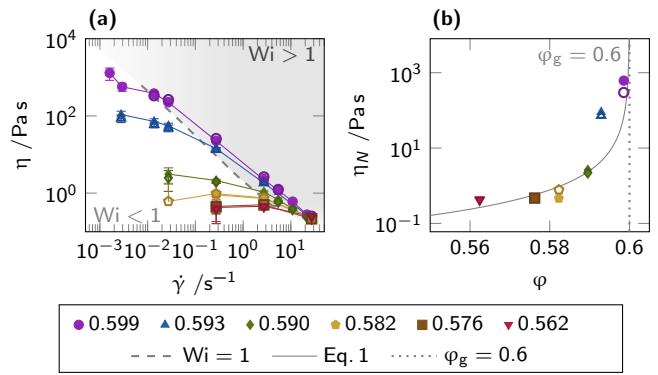
\begin{figure}[tb]
\centering
\begin{tikzpicture}
\sisetup{detect-all}
\normalfont\sffamily
\sansmath\sffamily

\def\xminval{5e-4}
\def\xmaxval{5e1}
\def\yminval{1e-1}
\def\ymaxval{1e4}

\begin{axis}[
          xlabel = {$\dot\gamma$~/\si{\per\second}},
		x label style={at={(rel axis cs:0.5,-0.37)},anchor=south},
	y label style={at={(rel axis cs:-0.19,0.5)},anchor=south},
		ylabel={$\eta$~/\si{\pascal\second }},
		xmode=log,	ymode=log,
          width=0.55\linewidth, height=0.48\linewidth,
          xmax=\xmaxval,
          xmin=\xminval,
          ymin=\yminval,
          ymax=\ymaxval,
          mark size=1.75pt,
          mark options={solid},
clip=false,
	xtick = {1e-4,1e-3,1e-2,1e-1,1e0,1e1,1e2},
	legend columns=6,
    legend style={
draw=none,		
only marks,
			font=\footnotesize\sffamily,
			fill=none,
		    legend cell align=left,
		    at={(1.1,-0.38)},
		    anchor=north,
		     /tikz/column 2/.style={column sep=6pt},
		     /tikz/column 4/.style={column sep=6pt},
		     /tikz/column 6/.style={column sep=6pt},
		     /tikz/column 8/.style={column sep=6pt},
		     /tikz/column 10/.style={column sep=6pt},
		     /tikz/column 12/.style={column sep=6pt},
		     /tikz/column 14/.style={column sep=6pt},          	
		}
    ]

\node[anchor=south west, xshift=-2pt] at (axis description cs: 0,1) {\sffamily \textbf{(a)}};
\node[anchor=south west, xshift=-2pt] at (axis description cs: 1.44,1) {\sffamily \textbf{(b)}};

\addplot [forget plot, fill=gray, fill opacity=0.2, draw=none]
coordinates{
(\xmaxval, 0.476924364)
(\xmaxval, \ymaxval)
(0.790220509, \ymaxval)
(0.790220509, 2.619782567)
(4.011099267, 0.476924364)
};

\addplot[name path=BagReg,  forget plot, 
draw=none, 
shade, left color=white, right color=gray,fill opacity=0.2,
shading angle=90,
]
coordinates
{
(0.790220509, \ymaxval)
(\xminval, \ymaxval)
(0.007416721, 616.3996759)
(0.040548795, 90.31362272)
(0.790220509, 2.619782567)
};

\addplot[name path=BagReg,  forget plot, 
draw=none, 
shade, left color=gray, right color=white,fill opacity=0.2,
shading angle=0,
]
coordinates
{
(\xmaxval, 0.476924364)
(4.011099267, 0.476924364)
(\xmaxval, \yminval)
};

\addplot [forget plot,gray,dashed,thick]
coordinates{
(0.007416721, 616.3996759)
(0.040548795, 90.31362272)
(0.790220509, 2.619782567)
(4.011099267, 0.476924364)
(7.266046324, 0.441156455)
};

\node [anchor=south west, gray] at (\xminval, \yminval) {\sffamily\footnotesize $\Pe <1$};
\node [anchor=north east, darkgray] at (\xmaxval, \ymaxval) {\sffamily\footnotesize $\Pe >1$};

\foreach \PF/\marque/\col/\variable/\extra in {
\PhiQfive/rottriangle*/c1/5_1/,
\PhiQthreepointfive/square*/c4/3.5_1/,
\PhiQthree/pentagon*/gold!80!c4/3_1/only marks, 
\PhiQtwopointfive/diamond*/c10!90!yellow/2.5_7/, 
\PhiQtwopointthree/triangle*/c22/2.3_1/, 
\PhiQtwo/*/c25/2_2/}{
    \edef\theoryplottemp{\noexpand\addplot [
    color=\col,
    mark=\marque,
	line legend,
    error bars/.cd, 
    	y dir=both, 
    	y explicit,
		error bar style={color=\col, solid}
    ]
    table[
restrict expr to domain={(\noexpand\thisrow{n_up}*\knval*2*pi)/60}{1e-3:30},
    	col sep=comma,
        x expr = (\noexpand\thisrow{n_up}*\knval*2*pi)/60,
        y expr = (\noexpand\thisrow{M_up}*1e-6/(2*pi*\lengthL*(\Di/2)^2))/((\noexpand\thisrow{n_up}*\knval*2*pi)/60),
        y error plus expr= (\noexpand\thisrow{stdM_up}*1e-6/(2*pi*\lengthL*(\Di/2)^2))/((\noexpand\thisrow{n_up}*\knval*2*pi)/60),
        y error minus expr=(\noexpand\thisrow{stdM_up}*1e-6/(2*pi*\lengthL*(\Di/2)^2))/((\noexpand\thisrow{n_up}*\knval*2*pi)/60),
    ]
	{Olfa_data/Q\variable.csv};
	\noexpand\addlegendentry{\round{3}{\PF} }
	}\theoryplottemp
}

\foreach \marque/\col/\variable in {
o/c25/2_2,
triangle/c22/2.3_1, 
diamond/c10!90!yellow/2.5_7, 
pentagon/gold!80!c4/3_1, 
square/c4/3.5_1,
rottriangle/c1/5_1}{
    \edef\theoryplottemp{\noexpand\addplot [
    forget plot,
    color=\col!80!black, thick, only marks,
    mark=\marque,
    error bars/.cd, 
    	y dir=both, 
    	y explicit,
		error bar style={color=\col!80!black, solid}
    ]
    table[
restrict expr to domain={(\noexpand\thisrow{n_down}*\knval*2*pi)/60}{1e-3:30},
        col sep=comma,
        x expr = (\noexpand\thisrow{n_down}*\knval*2*pi)/60,
        y expr = (\noexpand\thisrow{M_down}*1e-6/(2*pi*\lengthL*(\Di/2)^2))/((\noexpand\thisrow{n_down}*\knval*2*pi)/60),
        y error plus expr= (\noexpand\thisrow{stdM_down}*1e-6/(2*pi*\lengthL*(\Di/2)^2))/((\noexpand\thisrow{n_down}*\knval*2*pi)/60),
        y error minus expr=(\noexpand\thisrow{stdM_down}*1e-6/(2*pi*\lengthL*(\Di/2)^2))/((\noexpand\thisrow{n_down}*\knval*2*pi)/60),
    ]
{Olfa_data_divided/Q\variable-down.csv};
	}\theoryplottemp
}
\foreach \marque/\col/\variable in {
pentagon/gold!80!c4/3_1}{
    \edef\theoryplottemp{\noexpand\addplot [
    forget plot,
    color=\col, no marks,
    ]
    table[
restrict expr to domain={(\noexpand\thisrow{n_down}*\knval*2*pi)/60}{1e-3:30},
        col sep=comma,
        x expr = (\noexpand\thisrow{n_down}*\knval*2*pi)/60,
        y expr = (\noexpand\thisrow{M_down}*1e-6/(2*pi*\lengthL*(\Di/2)^2))/((\noexpand\thisrow{n_down}*\knval*2*pi)/60),
        y error plus expr= (\noexpand\thisrow{stdM_down}*1e-6/(2*pi*\lengthL*(\Di/2)^2))/((\noexpand\thisrow{n_down}*\knval*2*pi)/60),
        y error minus expr=(\noexpand\thisrow{stdM_down}*1e-6/(2*pi*\lengthL*(\Di/2)^2))/((\noexpand\thisrow{n_down}*\knval*2*pi)/60),
    ]
{Olfa_data_divided/Q\variable-down.csv};
	}\theoryplottemp
}

\end{axis}

\begin{axis}[
		xshift=0.528\linewidth,
	xlabel = {$\varphi$},
	x label style={at={(rel axis cs:0.5,-0.32)},anchor=south},
	y label style={at={(rel axis cs:-0.22,0.5)},anchor=south},
          ylabel={$\eta_N$~/\si{\pascal\second }},
		ymode=log,
          width=0.52\linewidth, height=0.48\linewidth,
          xmin=0.55,
          ymin=6e-2,
          ymax=8e3,
          mark size=1.75pt,
          mark options={solid},
	legend columns=6,
    legend style={
			font=\footnotesize\sffamily,
			fill=none,draw=none,
		    legend cell align=left,
		    at={(-0.4,-0.54)},
		    anchor=north,
		     /tikz/column 2/.style={column sep=6pt},
		     /tikz/column 4/.style={column sep=6pt},
		     /tikz/column 6/.style={column sep=6pt},
		     /tikz/column 8/.style={column sep=6pt},
		     /tikz/column 10/.style={column sep=6pt},
		     /tikz/column 12/.style={column sep=6pt},
		     /tikz/column 14/.style={column sep=6pt},
          	}
    ]

\addlegendimage{gray, dashed, thick, no marks}
\addlegendentry{$\Pe = 1$}  

\def\etaequation{1.5}
\def\phig{0.6}
\def\etazero{2e-3}

\addplot[
	darkgray, 
densely dashed, dash pattern=on 7pt off 2pt,dash phase=9pt,
	domain = 0.55:0.6,
	samples=500, 
] {
(\etazero)/((\phig-x)^(\etaequation))
};
\addlegendentry{Eq.\,\ref{eq:newtonianviscosity} (\acrshort{kd})}

\def\etaequation{2.4}
\def\phig{0.597}
\def\etazero{4e-5} 
\addplot[
	gray!80,
	domain = 0.55:0.6,
	samples=500, 
] {
(\etazero)/((\phig-x)^(\etaequation))
};
\addlegendentry{Eq.\,\ref{eq:newtonianviscosity} (\acrshort{mct})}

\foreach \n/\PF/\marque/\col/\variable in {
0/\PhiQtwo/*/c25/2_2,
1/\PhiQtwopointthree/triangle*/c22/2.3_1,
2/\PhiQtwopointfive/diamond*/c10!90!yellow/2.5_7, 
3/\PhiQthree/pentagon*/gold!80!c4/3_1, 
4/\PhiQthreepointfive/square*/c4/3.5_1,
5/\PhiQfive/rottriangle*/c1/5_1}{
    \edef\theoryplottemp{\noexpand\addplot [
    color=\col,
    mark=\marque,
    only marks, forget plot,
    line legend,
    error bars/.cd, 
    	y dir=both, 
    	y explicit
    ]
table [select coords between index={\n}{\n}, 
    	col sep=comma,	
		x expr=\PF,
		y expr=\noexpand\thisrow{Visc_up_Newtonian},
	] 
{NewtonianViscosities.data};
	}\theoryplottemp
}

\foreach \n/\PF/\marque/\col/\variable in {
0/\PhiQtwo/o/c25/2_2, 
1/\PhiQtwopointthree/triangle/c22/2.3_1, 
2/\PhiQtwopointfive/diamond/c10!90!yellow/2.5_7, 
3/\PhiQthree/pentagon/gold!80!c4/3_1, 
4/\PhiQthreepointfive/square/c4/3.5_1,
5/\PhiQfive/rottriangle/c1/5_1}{
    \edef\theoryplottemp{\noexpand\addplot [
    color=\col!80!black,
    mark=\marque,
    only marks, forget plot,
	thick,
    line legend,
    error bars/.cd, 
    	y dir=both, 
    	y explicit
    ]
table [select coords between index={\n}{\n}, 
    	col sep=comma,	
		x expr=\PF,
		y expr=\noexpand\thisrow{Visc_down_Newtonian},
	] 
{NewtonianViscosities.data};
	}\theoryplottemp
}

\def\phig{0.6}
\addplot    [
	dotted, thick, gray, 
	no marks,
]
coordinates{
(\phig, 6e-2)
(\phig, 8e3)
        };
\addlegendentry{$\varphi_\text{g}$}      

\node [anchor=south east, gray] at (\phig, 5e-2) {$\varphi_\text{g} =0.6$};

\end{axis}

\draw [] (rel axis cs: -0.23, -0.14) rectangle (rel axis cs:2.55, -.495);

\end{tikzpicture}
\caption{\label{fig:viscosity} Fluidized granular bed apparent
  viscosity, $\eta = \sigma / \dot\gamma $, in the Newtonian and shear
  thinning regimes. (a)~Viscosity, $\eta(\dot\gamma)$ \textit{vs.}\ shear rate,
  $\dot\gamma = K_\text{N}\varOmega$ (see text for
  details). P\'eclet number $\Pe = 1$ (\textit{cf.}\ Eq.\,\ref{eq:Wi}) is indicated
  by the dashed line, $\Pe > 1$ by shaded background. 
(b)~Newtonian viscosity, $\eta_N$, averaged over
  the relevant $\dot\gamma$ for each packing fraction, $\varphi$; the
  vertical dotted line indicates $\varphi_\mathrm{g}$. 
Dashed line indicates Eq.\,(\ref{eq:newtonianviscosity}) for the \gls{kd} relation; 
solid line Eq.\,(\ref{eq:newtonianviscosity}) according to \gls{mct} predictions.
}
\end{figure}

\textit{Newtonian rheology.---}Characterized by the linear relation
between shear stress and shear rate (or angular velocity),
$\sigma \propto \varOmega$, a Newtonian regime is observed at low
shear and density.
To translate the rotation rate, $\varOmega$, 
into the shear rate in the gap, $\dot\gamma = K \varOmega$, 
we use the well-known strain constant
$K_\mathrm{N} = 2 \delta^2 / (\delta^2-1)$ (where
$\delta = R_o / R_i$, ratio of outer to inner shear cell radii).
As the velocity gradient depends on both shear geometry and fluid's rheology, 
different rheological regimes imply different strain constants;
$K_\mathrm{N} $ encodes the Taylor-Couette geometry for Newtonian fluids (see Appendix II for full derivation).
Plotting $\eta(\dot\gamma) = \sigma / \dot\gamma $ (Fig.\,\ref{fig:viscosity}a) 
exhibits this regime of constant viscosity, $\eta_\mathrm{N} := \eta(\dot\gamma\to0)$.

In this regime, fluidization dominates over shear.
The dissipative collisions provide
an energy sink with a power density
$\Pi_\mathrm{c}(T_0)$ that balances the fluidization,
$\Pi_\mathrm{f} = \Pi_\mathrm{c}$, 
and fixes a constant granular temperature in the unsheared state, $T_0$.
$\Pi_{\dot\gamma} := \sigma\dot\gamma$, remains negligible compared to
the effect of fluidization, $\Pi_{\dot\gamma} \ll \Pi_\mathrm{f}$,
the granular temperature $T_0$ does not change
appreciably \cite{Kranz2018}.
Qualitatively, the rheology of the fluidized bed
is the same as that of colloidal suspensions.
From this analogy, Newtonian rheology is
expected, with our shear rate independent viscosity,
$\eta_\mathrm{N}$, increasing with density
\cite{Wagner2021book,Garcia2013,Kranz2010}.

As a function of packing fraction,
we indeed measure a strong increase in Newtonian viscosity, $\eta_\mathrm{N}(\varphi)$,
captured by a power law divergence (Fig.\,\ref{fig:viscosity}b),
\begin{equation}
  \label{eq:newtonianviscosity}
  \eta_\text{N}(\varphi) \propto (\Phi - \varphi)^{-\gamma}.
\end{equation}
For dense suspensions, 
Eq.\,(\ref{eq:newtonianviscosity}) is known as the \gls{kd}
relation \cite{Krieger1959},
where $\Phi = \varphi_\mathrm{g}$ denotes 
the maximal concentration of the suspension
at which $\eta_\text{N}$ diverges (here, $\varphi_\mathrm{g}=0.6$)
and $\gamma = 2.5 \varphi_\text{g} $ (dashed line in Fig.\,\ref{fig:viscosity}b).
\Gls{mct} can also be used to
predict the divergence of $\eta_\mathrm{N}$ upon approaching
a critical density, $\Phi = \varphi_\mathrm{c}$,
with $\varphi_\mathrm{c} < \varphi_\mathrm{g}$ \cite{Janssen2018,Kranz2010}.
The system-specific parameter $\gamma$ 
assumes $\gamma \approx 2.4$ \cite{Kranz2010}, and $\varphi_\mathrm{c} = 0.597$
(solid line in Fig.\,\ref{fig:viscosity}b).
Above $\varphi_\mathrm{c}$, the fluidized bed arrests into an amorphous
solid, with all particles still agitated ($T_0 > 0$), but unable to
move over long distances, due to the cage-like structure
formed by their neighbors \cite{Abate2006, Reis2007, Kranz2010}.

\textit{Dynamic yield stress.---} While a true yield stress in
suspensions has been subject to debate \cite{Barnes1999,Stickel2005},
an apparent yield stress at low $\dot\gamma$ and high $\varphi$ was
since measured in emulsions \cite{Negro2023}, foams
\cite{Katgert2009}, colloidal \cite{Pham2006} and granular suspensions
\cite{Fall2009}. At our highest packing fraction,
$\varphi = 0.599\lesssim\varphi_\mathrm{g}$, 
we observe a well-defined plateau in the flow
curves, from which we can extrapolate a finite dynamic yield stress,
$\sigma_0 := \sigma(\varphi_\mathrm{g}) \approx \SI{6}{\pascal}$, for
the emerging granular glass.

In non-agitated granular solids,
cohesive and frictional forces in lasting particle contacts result in a
\emph{static} yield stress \cite{Coussot1995,Kostynick2022,Pradeep2024}. 
By fluidization, contacts between particles are explicitly
broken, such that particle interactions cannot explain the
dynamic yield stress observed.
Such dynamic yield stress at the transition to an amorphous solid is expected
at the glass transition \cite{Pham2008,Fuchs2002,Kranz2018}. 
Due to the continuous agitation, 
particles stay in motion ($T_0 > 0$), even in the granular glass state.
This is incompatible with a static ($T_0 = 0$), jammed configuration, 
with lasting particle-particle contacts. It is however perfectly compatible with
a glassy state of matter, where local motion of a particle is
permitted, but long range motion is suppressed by the cage-like
structure formed by the particles' neighbors \cite{Abate2006,
Reis2007, Kranz2010}.

A $\varphi_\mathrm{g} = 0.6$ is
consistent with expectations. For colloidal suspensions in thermal
equilibrium, $\varphi_\text{g}\simeq0.57$-$0.58$ \cite{vanMegen1995}.
A higher $\varphi_\text{g}$ for our agitated granular system can be
attributed to its characteristics: (i) polydispersity (within
$\nicefrac{1}{4}\,d$) allows denser packing \cite{Gotze2003}, and (ii)
dissipative particle collision require a higher critical density to
solidify \cite{Kranz2010}.
Hence, we identify
$\varphi_g$ as the density at a granular glass transition,
distinct from a jamming transition.

\textit{Shear thinning rheology.---}At intermediate $\dot\gamma$, a shear thinning regime appears, 
where $\eta$ decreases with increasing $\dot\gamma$
(Fig.\,\ref{fig:viscosity}a).

In Brownian suspensions, shear thinning emerges when shear-induced
particle motion becomes relevant compared to thermally activated diffusion.
A finite diffusivity is related to a finite structural relaxation time, $\tau$, that
diverges with the viscosity, $\tau \sim \eta_\text{N}$.
In our non-Brownian but constantly agitated air-fluidized bed,
it is the imposed granular temperature that induces a time-scale, $\tau$,
competing with that of shear, $1/\dot\gamma$.
The P\'eclet number captures this competition:
$\Pe \ll 1$ is associated with slow, fluidization-induced flow and $\Pe\gg1$ with
shear dominated flow.
In the latter case, the fluid structure can no
longer instantaneously adapt to the imposed shear and effectively
behaves as a yield stress fluid, resulting in a shear thinning rheology.

We define a proxy for the characteristic fluidization-induced time-scale,
$\tau(\varphi) = \eta_\mathrm{N}(\varphi)/\sigma(\dot\gamma_0 | \varphi)$,
by generalizing the yield stress to a typical stress value,
$\sigma_0(\varphi) := \sigma(\dot\gamma_0  | \varphi)$ \cite{Coussot1995,Coussot1999,Kostynick2022,Pradeep2024}.
For the lower packing fractions, $\varphi \ll \varphi_\text{g}$, 
we use the stress at the flow curves' inflection point, $\sigma (\dot\gamma_0 )$,
at the end of the Newtonian regime. 
\footnote{Precisely, we define the flow curve's
inflection point, $\dot\gamma_0$, as $d^2\ln\sigma/d(\ln\dot\gamma)^2|_{\dot\gamma = \dot\gamma_0} = 0$.}.
This allows us to calculate the corresponding P\'eclet number
\footnote{We keep the notion of a P\'eclet number to stress the conceptual continuity 
between Brownian and non-Brownian agitated suspensions (see also e.g.\ \cite{Coussot1999,Hanotin2012});
however, if we interpret $\tau$ as a structural rather than a
diffusive time scale, one might be equally well justified to think
of $\Pe$ as a Weissenberg number \cite{White1964, Fuchs2002}.},
\begin{equation}\label{eq:Wi}
  \Pe := \eta_\mathrm{N}(\varphi)\dot\gamma/\sigma(\dot\gamma_0  | \varphi ).
\end{equation}

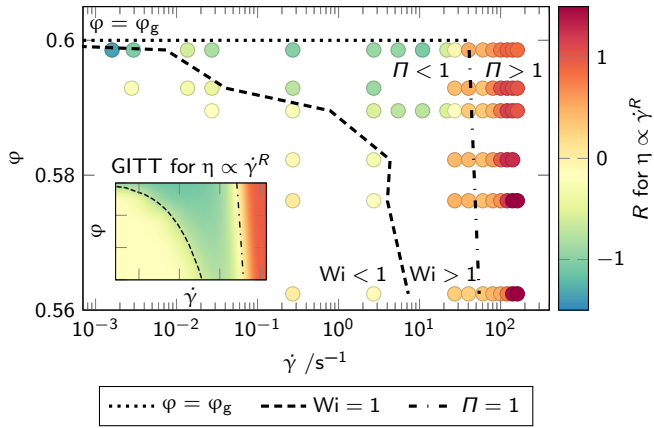
\begin{figure}[tb]
  \centering
  \begin{tikzpicture}
\sisetup{detect-all}
\normalfont\sffamily
\sansmath\sffamily

\def\minx{7e-4}
\def\maxx{4e2}
\def\miny{0.56}
\def\maxy{0.605}
\def\figwidth{0.9}
\def\figheight{0.65}

\begin{axis}[
	xmode=log,
          xlabel = {$\dot\gamma$~/\si{\per\second}},
          ylabel = {$\varphi$},
          width=\figwidth\linewidth, height=\figheight\linewidth,
          xmin=\minx,
          xmax=\maxx,
          ymin=\miny,
          ymax=\maxy,
          mark options={solid},
        colormap={Spectral_r}{
            rgb255=(50, 136, 189)   
            rgb255=(102, 194, 165)  
            rgb255=(171, 221, 164)  
            rgb255=(230, 245, 152)  
            rgb255=(255, 255, 191)  
            rgb255=(254, 224, 139)  
            rgb255=(253, 174, 97)   
            rgb255=(244, 109, 67)   
            rgb255=(213, 62, 79)    
            rgb255=(158, 1, 66)     
        },
    colorbar,
    colorbar style={
        ylabel={$R$ for $\eta \propto \dot{\gamma}^R$},
		width=0.047*\axisdefaultwidth,
            at={(1.02,0.5)},
            anchor=west,
        	ylabel style={yshift=0.5em},
    },
    every axis plot/.append style={mesh, solid, point meta=explicit},
    point meta min=-1.5,
    point meta max=1.5,
    ]

\foreach \PF/\marque/\col/\variable/\m/\n in {
\PhiQtwo/*/c25/2_2/0/4,
\PhiQtwopointthree/triangle*/c22/2.3_1/0/4,
\PhiQtwopointfive/diamond*/c10!90!yellow/2.5_7/0/3, 
\PhiQthree/pentagon*/gold!80!c4/3_1/0/1,
\PhiQthreepointfive/square*/c4/3.5_1/0/1,
\PhiQfive/rottriangle*/c1/5_1/0/1}{
    \edef\theoryplottemp{\noexpand\addplot [
    mark=*,
	mark size = 2.7pt,
    only marks, forget plot,
	scatter, scatter src=explicit, 
    ]
    table[
    	col sep=comma,
x index=2,y index=1,meta index=0,
    ]
		{Olfa_data_divided/calcQ\variable.txt};
	}\theoryplottemp
}

\coordinate (colormappos) at (rel axis cs:1.02,1);

\end{axis}

\begin{axis}[
	xmode=log,
          xlabel = {},
          ylabel = {},
title style={at={(rel axis cs:0.5,0.9)},anchor=south},
          width=\figwidth\linewidth, height=\figheight\linewidth,
          xmin=\minx,
          xmax=\maxx,
          ymin=\miny,
          ymax=\maxy,
	xticklabels ={},
	yticklabels ={},
xlabel={},
ylabel={},
	area legend,
	legend columns=3,
    legend style={
			font=\footnotesize\sffamily,
			fill=none,
		    legend cell align=center,
		    at={(0.5,-0.25)},
		    anchor=north,
		     /tikz/column 2/.style={column sep=8pt},
		     /tikz/column 4/.style={column sep=8pt},
          	}
    ]

\def\phig{0.6}

\addplot [
	color=black,
	dotted, very thick,
    line legend,
    ]
coordinates{
(\minx, \phig)
(4e1, \phig)
};
\addlegendentry{$\varphi = \varphi_\text{g}$}  
\node [anchor=south west, xshift=-1pt,yshift=-1pt] at (\minx, \phig) {\sffamily\footnotesize $\varphi = \varphi_\text{g}$};

\addplot [
	color=black,
	densely dashed, very thick,
    line legend,
    ]
    table[
    	col sep=tab,
        x expr = \thisrow{gammadot1}, 
        y expr = \thisrow{Phi},
    ]
{NewtonianThinning.data};
\addlegendentry{$\Pe = 1$}
\node [anchor=south east, ] at (5e0,0.563) {\sffamily\footnotesize $\Pe <1$};
\node [anchor=south west, ] at (6e0,0.563) {\sffamily\footnotesize$\Pe > 1$};

\addplot [
	color=black,
	loosely dash dot, very thick,
    line legend,
    ]
    table[
    	col sep=tab,
        x expr = \thisrow{gammadot2}, 
        y expr = \thisrow{Phi},
    ]
{ShearThinningBagnold.data};
\addlegendentry{$\Pi = 1$}  
\node [anchor=north east, ] at (3e1,0.598) {\sffamily\footnotesize $\Pi <1$};
\node [anchor=north west, ] at (5e1,0.598) {\sffamily\footnotesize $\Pi >1$};

\coordinate (insetPosition) at (rel axis cs:0.07,0.1);

\end{axis}

\begin{axis}[
font = \footnotesize,
at={(insetPosition)},anchor={south west},
width=0.2\textwidth, height=0.16\textwidth, 
xlabel={$\dot{\gamma}$},
ylabel={$\varphi$},
x label style={at={(rel axis cs:0.5,0)} },
y label style={at={(rel axis cs:-0.04,0.5)} },
xmode=log,
xticklabels ={},
yticklabels ={},
enlarge x limits =0,
enlarge y limits =0,
xmode = log,
	ymin=0.5,
	ymax=0.53,
	xmin=7e-9,
	xmax=2e-1,
]
\addplot graphics [
	ymin=0.5,
	ymax=0.53,
	xmin=7e-9,
	xmax=2e-1,
] {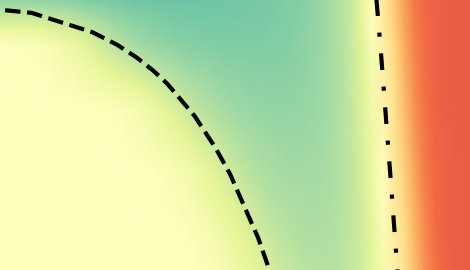};
\coordinate (colormappos) at (rel axis cs:1,1);

\end{axis}

\begin{axis}[
	font = \footnotesize, 
at={(insetPosition)},anchor={south west},
width=0.2\textwidth, height=0.16\textwidth, 
title ={GITT for $\eta \propto \dot{\gamma}^R$},
title style={at={(rel axis cs:0.5,0.82)},anchor=south},
xticklabels ={},
yticklabels ={},
xlabel={},
ylabel={},
xtick={1e-7, 1e-5, 1e-3,1e-1},
ytick={},
xmode = log,
	ymin=0.5,
	ymax=0.53,
	xmin=7e-9,
	xmax=2e-1,
]

\end{axis}

\end{tikzpicture}
  \caption{\label{fig:energybalance} Rheological state diagram spanned
    by packing fraction $\varphi$ and shear rate $\dot\gamma$. The
    flow index $R$ ($\eta \propto \dot\gamma^R$) is color coded, where
    $R=0$ corresponds to Newtonian rheology and $R < 0$ ($R> 0$)
    indicates shear thinning (thickening) behavior. The dashed
    line traces P\'eclet number $\Pe=1$, and the dash-dotted
    line delineates power ratio $\Pi=1$. The dotted line marks the
    granular glass transition at $\varphi_\mathrm{g}=0.6$.
}
\end{figure}

In Figs.\,\ref{fig:viscosity}a and \ref{fig:energybalance} we find,
indeed, $\Pe \sim 1$
to trace the crossover from Newtonian to shear 
thinning behavior,
generic for dense suspensions \cite{Stickel2005,Fuchs2002,Brader2009}.
As long as shear-induced agitation is negligible, 
we find the analogy to colloidal suspensions to still hold.

\begin{figure}[tb]
\centering
\input{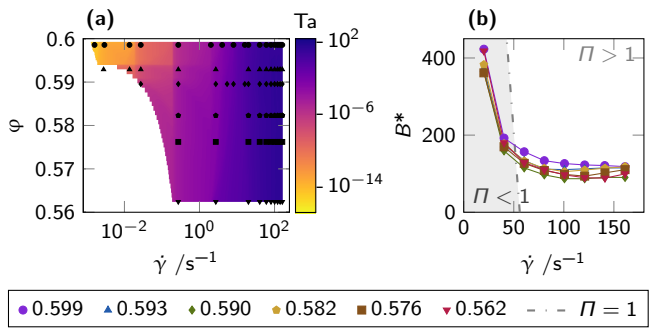}
\caption{\label{fig:exp-Bagnold} Flow curves analysis for the high
  shear rate regime (where $\dot\gamma = K_\text{B}\varOmega$,
  see text for details).  (a)~Taylor number, $\Ta$, calculated on the
  full range of experimental data available. Location of measurements
  indicated by black dots.  (b)~Dimensionless Bagnold coefficients,
  $B$* as a function of shear rate $\dot\gamma$ for the packing
  fraction as given in the legend. The dash-dotted line displays the
  location of power ratio $\Pi=1$, marking the onset of the Bagnold
  regime }
\end{figure}

\textit{Bagnoldian rheology.---}The shear thickening regime appears once
$\varOmega \gtrsim \SI{10}{\per\second}$, and exhibits
$\sigma \sim \varOmega^2 $ (see Fig.\,\ref{fig:experimental}), a scaling
first identified by Bagnold in granular suspension \cite{Bagnold1954}.
As the material's flow profile varies with its rheology, $\dot\gamma$
is related to the angular velocity, $\varOmega$, by a different strain
constant, $ K_\text{B} = \delta / (\delta - 1) $ (full derivation in Appendix~II), 
\red{that captures the specifics of our shear geometry, 
only now for a Bagnoldian instead of a classical Newtonian fluid.}
The data in Fig.\ \ref{fig:exp-Bagnold} is
plotted for $\dot\gamma = K_\text{B}\varOmega$.

First, we want to assess whether 
Taylor vortices,
previously observed in granular fluidized beds \cite{Conway2004}, 
and which could explain a significant increase in $\sigma$ \cite{Khali2013},
might appear in our system.
We calculate the gap Reynolds and Taylor numbers for the granular fluidized bed.
The former, $\Rey = \rho_b \varphi R_i \varOmega (R_o-R_i) / \eta$,
compares the time-scale of rotational advection to that of viscous damping.
It remains small throughout the experiment ($\Rey < 10$) (details in Appendix III),
suggesting laminar flow \cite{Coussot1999,Ramesh2019,Dash2020,Majji2018, Baroudi2023}.
The Taylor number,
$\Ta$ \footnote{We use the definition of \citeauthor{DiPrima1984}
\cite{DiPrima1984}, $\Ta = \kappa \Rey^2$, where the prefactor
$\kappa = 2(\delta - 1)/(\delta + 1)$ is related to the shear geometry.},
compares Coriolis to viscous forces \cite{DiPrima1984,Dash2020}.
Taylor vortices are expected to emerge if $\Ta$ exceeds a critical value,
estimated 
around $\Ta_\mathrm{c} \sim \orderof (10^3)$ or above
\cite{DiPrima1984,Dash2020,Gillissen2019}
(details in Appendix III). 
Fig.\,\ref{fig:exp-Bagnold}a shows that we do not achieve
Taylor numbers higher than $\Ta \lesssim 10^2\ll\Ta_\mathrm{c}$, such that
shear thickening cannot be explained by the Taylor instability.

Dense suspensions generally feature a shear thickening regime at high
shear rates \cite{Stickel2005, Brown2014, Lemaitre2009}.
In Brownian suspensions,
two primary mechanisms are associated with shear thickening: 
hydrodynamic effects and the lubrication-to-friction transition \cite{Brown2014,Lin2015,Morris2018,Jamali2019}.
But the shear thickening observed in the Bagnold regime has a different origin.
While in Brownian suspensions, the interstitial fluid
is in thermal equilibrium with the particles and can absorb the effect of
shear heating, in the fluidized bed, the granular
temperature is decoupled from the air's thermodynamic temperature.
The granular temperature is hence immediately increased by shear heating, resulting in an
increased shear stress.

The crossover to the Bagnold regime is
expected once shear heating, $\Pi_{\dot\gamma}$, becomes comparable to
the fluidization power \red{density}, $\Pi_\mathrm{f}$, i.e., at a ratio
\begin{equation}
  \label{eq:2}
  \Pi := \Pi_{\dot\gamma}/\Pi_\mathrm{f}
  = \sigma\dot\gamma/\rho_p\varphi ug
\end{equation}
of the order of one. In Figs.\ \ref{fig:exp-Bagnold}b and
\ref{fig:energybalance}, we find that the power density ratio, $\Pi$, indeed controls
the emergence of the Bagnold behavior.

\begin{figure}[tb]
	\centering 
	\begin{tikzpicture}
\sisetup{detect-all}
\normalfont\sffamily
\sansmath\sffamily

\def\ymin{\pgfkeysvalueof{/pgfplots/ymin}}
\def\ymax{\pgfkeysvalueof{/pgfplots/ymax}}
\def\xmin{\pgfkeysvalueof{/pgfplots/xmin}}
\def\xmax{\pgfkeysvalueof{/pgfplots/xmax}}

\begin{loglogaxis}[
          xlabel = {$\dot{\gamma}$~/\si{\per\second}},
          ylabel={$\sigma$~/\si{\pascal}},
	x label style={at={(rel axis cs:0.5,-0.11)},anchor=north},
	y label style={at={(rel axis cs:-0.13,0.5)} },
          width=0.7\linewidth, height=0.57\linewidth,
          xmin=1e-4,
          xmax=4e2,
          ymin=1e-2,
          ymax=5e2,
		ytick distance=1e1,
		xtick distance=1e1,
          mark size=1.75pt,
          mark options={solid},
	legend columns=3,
    legend style={
			only marks,
			font=\footnotesize\sffamily,
			fill=none,
		    legend cell align=left,
		    at={(0.09,0.985)},
		    anchor=north west,
		     /tikz/column 2/.style={column sep=5pt},
		     /tikz/column 4/.style={column sep=5pt},
          	}
    ]

\node [anchor=south east] at (4e2, 1e-2) {\sffamily\footnotesize $\varepsilon = 0.8$};


\foreach \marque/\col/\variable in {
*/c25/2.0,
triangle*/c22/2.3, 
diamond*/c10!90!yellow/2.5, 
pentagon*/gold!80!c4/3, 
square*/c4/3.5,
rottriangle*/c1/5}{
    \edef\theoryplottemp{\noexpand\addplot [
    color=\col!90!black,
    mark=\marque,
    no marks,
	forget plot,
    line legend,
    error bars/.cd, 
    	y dir=both, 
    	y explicit
    ]
    table[
    	col sep=tab,
x expr = \noexpand\thisrow{U}*\knval*2*pi,
y expr = \noexpand\thisrow{sigma},
    ]
	{Till_data_epsilon=0.8/GITTslow\variable.data}; 
	}\theoryplottemp
}

\foreach \marque/\col/\variable in {
o/c25/2.0,
triangle/c22/2.3, 
diamond/c10!90!yellow/2.5,
pentagon/gold!80!c4/3, 
square/c4/3.5,
rottriangle/c1/5}{
    \edef\theoryplottemp{\noexpand\addplot [
    forget plot,
    color=\col!80!black,
    mark=\marque,
    no marks,
    error bars/.cd, 
    	y dir=both, 
    	y explicit
    ]
    table[
        col sep=tab,
x expr = \noexpand\thisrow{U}*\kbval*2*pi,
y expr = \noexpand\thisrow{sigma},
    ]
	{Till_data_epsilon=0.8/GITTfast\variable.data}; 
	}\theoryplottemp
}


\foreach \PF/\marque/\col/\variable/\m/\n in {
\PhiQtwo/*/c25/2_2/0/4,
\PhiQtwopointthree/triangle*/c22/2.3_1/0/4,
\PhiQtwopointfive/diamond*/c10!90!yellow/2.5_7/0/3, 
\PhiQthree/pentagon*/gold!80!c4/3_1/0/1,
\PhiQthreepointfive/square*/c4/3.5_1/0/1,
\PhiQfive/rottriangle*/c1/5_1/0/1}{
    \edef\theoryplottemp{\noexpand\addplot [
    color=\col,
    mark=\marque,
    only marks,
	forget plot,
    line legend,
    error bars/.cd, 
    	y dir=both, 
    	y explicit
    ]
    table[select coords between index={\m}{\n}, 
    	col sep=comma,
        x expr = (\noexpand\thisrow{n_up}/60)*\knval*2*pi, 
        y expr = (\noexpand\thisrow{M_up}*1e-6)/(2*pi*\lengthL*(\Di/2)^2),
        y error plus expr=(\noexpand\thisrow{stdM_up}*1e-6)/(2*pi*\lengthL*(\Di/2)^2),
        y error minus expr=(\noexpand\thisrow{stdM_up}*1e-6)/(2*pi*\lengthL*(\Di/2)^2),
    ]
{Olfa_data_divided/Q\variable-up.csv};
	}\theoryplottemp
}

\foreach \PF/\marque/\col/\variable/\m/\n in {
\PhiQtwo/*/c25/2_2/5/12,
\PhiQtwopointthree/triangle*/c22/2.3_1/5/12, 
\PhiQtwopointfive/diamond*/c10!90!yellow/2.5_7/4/12, 
\PhiQthree/pentagon*/gold!80!c4/3_1/2/12, 
\PhiQthreepointfive/square*/c4/3.5_1/2/12,
\PhiQfive/rottriangle*/c1/5_1/2/12}{
    \edef\theoryplottemp{\noexpand\addplot [
    color=\col,
    mark=\marque,
    only marks,
    error bars/.cd, 
    	y dir=both, 
    	y explicit
    ]
    table[select coords between index={\m}{\n}, 
    	col sep=comma,
        x expr = (\noexpand\thisrow{n_up}/60)*\kbval*2*pi, 
        y expr = (\noexpand\thisrow{M_up}*1e-6)/(2*pi*\lengthL*(\Di/2)^2),
        y error plus expr=(\noexpand\thisrow{stdM_up}*1e-6)/(2*pi*\lengthL*(\Di/2)^2),
        y error minus expr=(\noexpand\thisrow{stdM_up}*1e-6)/(2*pi*\lengthL*(\Di/2)^2)
    ]
	{Olfa_data_divided/Q\variable-up.csv};
	\noexpand\addlegendentry{\round{3}{\PF} }
	}\theoryplottemp
}

\node[anchor=north west, xshift=-3pt, yshift=2pt] at (axis description cs: 0,1) {\sffamily \textbf{(a)}};
\end{loglogaxis}

\def\widthsmallgraphs{0.37}
\def\hsmallgraphs{0.3}
\def\shiftinx{0.67}
\def\shiftiny{0.135}

\begin{axis}[
xshift=\shiftinx\linewidth,
yshift=2*\shiftiny\linewidth,
          xlabel = {},
          ylabel={$\varphi_\text{th}$},
          width=\widthsmallgraphs\linewidth, height=\hsmallgraphs\linewidth,
          ymin=0.38,
          ymax=0.55,
	xticklabels = {,},
          mark size=1.75pt,
          mark options={solid},
	area legend,
	legend columns=5,
    legend style={
		    legend cell align=left,
		    at={(0.51,-0.21)},
		    anchor=north,
		     /tikz/column 2/.style={column sep=14pt}
          	}
    ]

\foreach \PF/\marque/\col/\variable/\m/\n in {
\PhiQtwo/*/c25/2_2/0/0,
\PhiQtwopointthree/triangle*/c22/2.3_1/1/1,
\PhiQtwopointfive/diamond*/c10!90!yellow/2.5_7/2/2, 
\PhiQthree/pentagon*/gold!80!c4/3_1/3/3,
\PhiQthreepointfive/square*/c4/3.5_1/4/4,
\PhiQfive/rottriangle*/c1/5_1/5/5}{
    \edef\theoryplottemp{\noexpand\addplot [
    color=\col,
    mark=\marque,
    only marks,
	forget plot,
    ]
    table[select coords between index={\m}{\n}, 
    	col sep=tab,
		x expr = \PF,
        y expr = \noexpand\thisrow{phi},
    ]
{Till_data_epsilon=0.8/GITTfitsU-short0.8.data};
	}\theoryplottemp
}

\node[anchor=north west, xshift=-2pt, yshift=2pt] at (axis description cs: 0,1) {\sffamily \textbf{(b)}};
\end{axis}

\begin{axis}[
xshift=\shiftinx\linewidth,
yshift=\shiftiny\linewidth,
          xlabel = {},
          ylabel={$P_0$/\,\si{\pascal}},
          width=\widthsmallgraphs\linewidth, height=\hsmallgraphs\linewidth,
          ymin=-0.05,
          ymax=1.25,
	xticklabels = {,},
          mark size=1.5pt,
          mark options={solid},
	area legend,
	legend columns=5,
    legend style={
		    legend cell align=left,
		    at={(0.5,-0.21)},
		    anchor=north,
		     /tikz/column 2/.style={column sep=14pt}
          	}
    ]

\foreach \PF/\marque/\col/\variable/\m/\n in {
\PhiQtwo/*/c25/2_2/0/0,
\PhiQtwopointthree/triangle*/c22/2.3_1/1/1,
\PhiQtwopointfive/diamond*/c10!90!yellow/2.5_7/2/2, 
\PhiQthree/pentagon*/gold!80!c4/3_1/3/3,
\PhiQthreepointfive/square*/c4/3.5_1/4/4,
\PhiQfive/rottriangle*/c1/5_1/5/5}{
    \edef\theoryplottemp{\noexpand\addplot [
    color=\col,
    mark=\marque,
    only marks,
	forget plot,
    ]
    table[select coords between index={\m}{\n}, 
    	col sep=tab,
	x expr = \PF,
	y expr = \noexpand\thisrow{nT0},
    ]
{Till_data_epsilon=0.8/GITTfitsU-short0.8.data};
	}\theoryplottemp
}

\node[anchor=north west, xshift=-2pt, yshift=2pt] at (axis description cs: 0,1) {\sffamily \textbf{(c)}};

\end{axis}

\begin{axis}[
xshift=\shiftinx\linewidth,
          xlabel = {$\varphi$},
          ylabel={$\omega_0$/\,\si{\kilo\hertz}},
          width=\widthsmallgraphs\linewidth, height=\hsmallgraphs\linewidth,
	scaled x ticks = false,
	x tick label style={
		/pgf/number format/precision=4
		},
clip=false,
          ymin=-0.5,
          ymax=11.5,
		xtick={0.56, 0.58, 0.6},
          mark size=1.5pt,
          mark options={solid},
	area legend,
	legend columns=5,
    legend style={
		    legend cell align=left,
		    at={(0.5,-0.21)},
		    anchor=north,
		     /tikz/column 2/.style={column sep=14pt}
          	}
    ]

\foreach \PF/\marque/\col/\variable/\m/\n in {
\PhiQtwo/*/c25/2_2/0/0,
\PhiQtwopointthree/triangle*/c22/2.3_1/1/1,
\PhiQtwopointfive/diamond*/c10!90!yellow/2.5_7/2/2, 
\PhiQthree/pentagon*/gold!80!c4/3_1/3/3,
\PhiQthreepointfive/square*/c4/3.5_1/4/4,
\PhiQfive/rottriangle*/c1/5_1/5/5}{
    \edef\theoryplottemp{\noexpand\addplot [
    color=\col,
    mark=\marque,
    only marks,
	forget plot,
    ]
    table[select coords between index={\m}{\n}, 
    	col sep=tab,
	x expr = \PF,
	y expr = \noexpand\thisrow{omega}/1000,
    ]
{Till_data_epsilon=0.8/GITTfitsU-short0.8.data};
	}\theoryplottemp
}

\node[anchor=north west, xshift=-2pt, yshift=2pt] at (axis description cs: 0,1) {\sffamily \textbf{(d)}};

\end{axis}

\end{tikzpicture}
	\caption{\label{fig:constitutiveequation} 
(a) Constitutive relation 
(Eq.\,\ref{eq:constitutiveequation}, solid lines) fitted to the
experimental data (marks) in the Newtonian and Bagnold regime.
The fit parameters (for coefficient of restitution
$\varepsilon=0.8$) are the theoretical packing fraction, $\varphi_\text{th}$,
ideal particle pressure in the unsheared fluid, $P_0$, and collision
frequency, $\omega_c$. They are given as a function of the experimental $\varphi$ in (b), (c), and (d), respectively.  }
\end{figure}
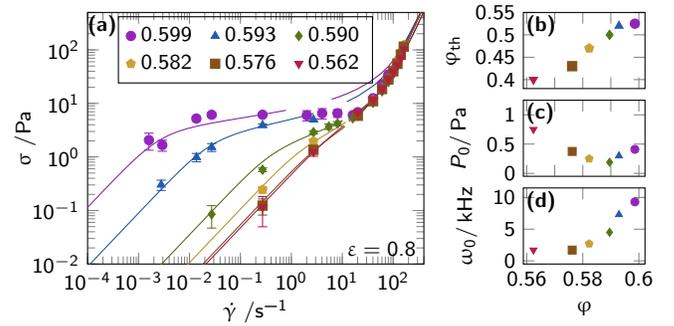

Remark also that shear stress collapses for all $\varphi$ once reaching the shear thickening regime (Figs.\  \ref{fig:experimental}, \ref{fig:viscosity}a, \ref{fig:exp-Bagnold}b, \ref{fig:constitutiveequation}a).
We attribute this to shear dominating in this regime:
fluidization-induced variations in packing density become negligible.
The high $\dot\gamma$ regime is hence indeed Bagnold shear thickening,
where $\dot\gamma$ becomes the only relevant time-scale in the system.

The Bagnold regime and its emergence, 
specific to granular materials \cite{Lemaitre2002}, 
have been studied in a number of
contexts~\cite{Forterre2008, Madraki2020, Tapia2022}; yet,
quantitative measurements of Bagnold coefficients are limited.  In
Fig.~\ref{fig:exp-Bagnold}b, we presents the Bagnold coefficients,
$B= \sigma/{\dot\gamma}^2$, expressed in dimensionless form, as
$B^* = Bd/m$ ($d$ and $m$ average particle diameter and mass).  We
find $B$* $\sim\orderof (100)$ 
(averaged over the four highest shear rates)
with no clear density dependence.
We compare our results to Bagnold's seminal measurements \cite{Bagnold1954}, 
where $0.1 < B$*$ < 10$, increasing with $\varphi$.
The suspension used in \cite{Bagnold1954} (wax spheres in liquids)
shows higher dissipation, consistent with the smaller $B$*.
compared to glass beads in air.

At this point, let us summarize that the flow curves of our
air-fluidized bed can be qualitatively characterized by two
dimensionless numbers: the power ratio, $\Pi$, and the P\'eclet
number, $\Pe$ (\textit{cf.}~Fig.~\ref{fig:energybalance}).  For $\Pi>1$, the
granular medium becomes purely shear driven---fluidization is
negligible---and Bagnold rheology applies. For $\Pi<1$, fluidization
controls the granular temperature $T_0$---shear is
negligible---and, in complete analogy to Brownian suspensions, the
rheology evolves with the \red{P\'eclet number} from Newtonian
($\Pe \ll 1$) to shear thinning ($\Pe \gg 1$).

\textit{Constitutive relation.---}The granular extension of the \gls{itt} formalism, \gls{gitt}~\cite{Kranz2018, Kranz2020},
takes this scaling analysis into account to extract 
the divergent relaxation time, $\tau$, as well as the power balance,
to provide a dimensionless constitutive relation for dissipative smooth hard sphere.

Details about \gls{gitt} are given in Appendix IV and elsewhere~\cite{Kranz2018, Kranz2020}.
The physical intuition behind this model is that of all stress relaxation modes, 
the slowest one in dense fluids will be the density fluctuations
\cite{Fuchs2002}. 
In the \gls{gitt} model, the dimensionless shear stress,
$\sigma(\dot\gamma/\omega_0\mid\varphi_\mathrm{th})/P_0$,
[Eq.~(\ref{eq:constitutiveequation})] is parameterized by the packing
fraction, $\varphi_\mathrm{th}$, and uses properties of the unsheared
fluid, namely, the collision frequency, $\omega_0$, and the ideal
particle pressure, $P_0$, as rate and stress scales, respectively.
\Gls{gitt} predicts the rheological state diagram presented as
inset of Fig.\,\ref{fig:energybalance}.

Given the qualitative similarity exhibited in Fig.\ \ref{fig:energybalance},
it is tempting to use the \gls{gitt} relation as a constitutive
model and fix its parameters by fitting. Note two things: firstly, we
do not know the shear rate, $\dot\gamma(\varOmega)$, in the shear thinning
regime. To determine the non-linear mapping that applies there is
beyond the scope of this letter \cite{DAngelo2024}, and we simply exclude the shear
thinning regime from the fit. Secondly, although we measured the packing
fraction, $\varphi$, we treat the packing fraction as a third
fit parameter, $\varphi_\mathrm{th}$. The glassy dynamics 
model, namely granular \gls{mct} \cite{Kranz2010}, at the heart
of \gls{gitt}, is known to produce a finite offset,
$\varphi - \varphi_\mathrm{th} > 0$, to the experimental values
of $\varphi$. This offset has hitherto not been quantified for granular fluidized beds.

In Fig.\,\ref{fig:constitutiveequation}a we present the result of a
manual fitting of the \gls{gitt} constitutive model to our
experimental flow curves.
A detailed
analysis of the fit parameters (Figs.\,\ref{fig:constitutiveequation}b,\,c,\,d) is left for future work, 
but let us note that they all assume reasonable values and depend smoothly on
the experimentally measured packing fraction, $\varphi$.
The ability of the \gls{gitt} constitutive model to capture both the Newtonian and Bagnold regimes, 
separated by many orders of magnitude in shear rate,
shows that the \acrshort{itt} formalism extends to off-equilibrium dynamics, and
further supports our scaling analysis of rheological regimes in air-fluidized granular beds, 
constituting the core of this contribution.

\textit{Conclusion.---}We measure the rheology of a granular bed, agitated by air-fluidization,
spanning five orders of magnitude in shear rate, $\dot\gamma$.
We capture a diverse rheology: Newtonian behavior at low $\dot\gamma$ and
packing density, $\varphi$; the development of an apparent dynamic yield stress around
$\varphi_\text{g}=0.6$, which we interpret as a granular glass transition;
Bagnoldian shear thickening at high $\dot\gamma$, 
where $\sigma$ collapses for all $\varphi$. 
The transitions between these regimes are
characterized by two dimensionless numbers: the shear-to-fluidization
power density ratio, $\Pi$, and the P\'eclet number, $\Pe$. 
While $\Pi < 1$, the granular bed behaves akin to Brownian suspensions,
$\Pe = 1$ marking the shift from Newtonian to shear thinning
behavior. At $\Pi > 1$, the material enters Bagnold regime, behaving
like sheared, unfluidized granular media. 
These different regimes are
qualitatively and quantitatively described by \gls{gitt}, using
underlying similarities between glasses, colloids, and granular
matter to propose a unified approach to understanding the rheology of
amorphous materials across scales.

With this work, we provide a framework to quantitatively characterize granular fluid's flow, on the same level as has been available for ordinary fluids. This will make granular flows amenable to continuum modeling in, e.g., industrial process design, geophysical hazard assessment, or predictions for environments challenging to access (e.g., for space exploration). 
More broadly, the approach presented here, relying on the importance of a glass transition and power balance,
might apply more generally to non-equilibrium fluids, notably active and biological matter---a step towards
a unified theoretical framework applicable from biological to astronomical systems.

\textit{Data availability statement.---}
The data that support the findings of this article are openly available \cite{Repository}.
The data shown in Fig.\,\ref{fig:experimental} was initially published in 
the doctoral thesis of O.\,D'Angelo
(Ref.\,\cite{DAngeloPhDthesis2021}).

\textit{Acknowledgements.---}
We are indebted to Dennis Schütz for 
sharing unpublished measurements that were the starting
point of this study. 
We thank Miriam Siebenb\"{u}rger  and Abhishek Shetty 
for discussions on the experimental procedures,
and Matthias Fuchs, Annette Zippelius and Olivier Coquand 
for companionship in the long development of
\gls{gitt} and many more interesting discussions. We thank Thomas
Voigtmann for sharing rheological insight and steady encouragement.


\bibliography{bibliography.bib}

\onecolumngrid

\newpage
\part{}
\twocolumngrid

\appendix
\renewcommand{\thefigure}{A\arabic{figure}}
\renewcommand{\theequation}{A\arabic{equation}}
\setcounter{figure}{0}  
\setcounter{equation}{0}  

\section{Appendix I:~Methods}

\begin{figure}[h!]
\centering
\begin{tikzpicture}
\sisetup{detect-all}
\normalfont\sffamily
\sansmath\sffamily
\footnotesize

\begin{scope}[scale=0.85, local bounding box=scope1, font=\footnotesize]

\draw[, thick, white] (9, 0.97)rectangle(13,3.6);

\draw[lightgray, thick, fill=lightgray!30] (9.97, 0.97)rectangle(12.027,3.6);

\fill[lightgray, thick] (10,1)rectangle(12,3.1);
\draw[black, thick] (10,1)rectangle(12,3.1);

\fill [gray] (11.5, 1.4)--(11.5, 3)--(11.0, 3.5)--(10.5, 3)--(10.5, 1.4)--cycle;
\fill[gray] (10.9, 3.3) rectangle (11.1, 3.9);

\draw [gray, thick, rounded corners=0.5mm, line join = round] (10.91, 4.1)--(10.95, 4.15)--(11.05, 4.05)--(11.09, 4.1);
\fill [gray, line join = round] (10.9, 3.8)--(10.9, 4.11)--(10.95, 4.126)--(11.05, 4.06)--(11.1, 4.09)--(11.1, 3.8)--cycle;

\draw [black, dash dot] (11, 0.8)--(11, 4.19);
\draw [Stealth-Stealth,black, ] (12.4, 1.4)--(12.4, 3)node[midway, anchor=west]{$L=38\,$\si{\mm}};
\draw [ dashed] (11.5, 1.4)--(12.4, 1.4);
\draw [ dashed] (11.5, 3)--(12.4, 3);

\draw [ dashed] (12, 1)--(12.4, 1);
\draw [Stealth-Stealth,black, ] (12.4, 1.4)--(12.4, 1)node[midway, anchor=west]{$10\,$\si{\mm}};

\draw [black, dashed] (10.5, 1.4)--(10.5, 0.77);
\draw [black, dashed] (11.5, 1.4)--(11.5, 0.77); 
\draw [Stealth-Stealth,black, ] (10.5, 0.77)--(11.5, 0.77)node[midway, anchor=north]{$D_i=24\,$\si{\mm}};

\draw [black, dashed] (10, 1.4)--(10, 0.2);
\draw [black, dashed] (12, 1.4)--(12, 0.2); 
\draw [Stealth-Stealth,black, ] (10, 0.2)--(12, 0.2)node[midway, anchor=north]{$D_o=50\,$\si{\mm}};

\end{scope}

\begin{scope}[shift={($(scope1.north west)+(0.2,-0.1)$)}, anchor=north east]
\node[] (cylinder) {
\includegraphics[height=90pt]{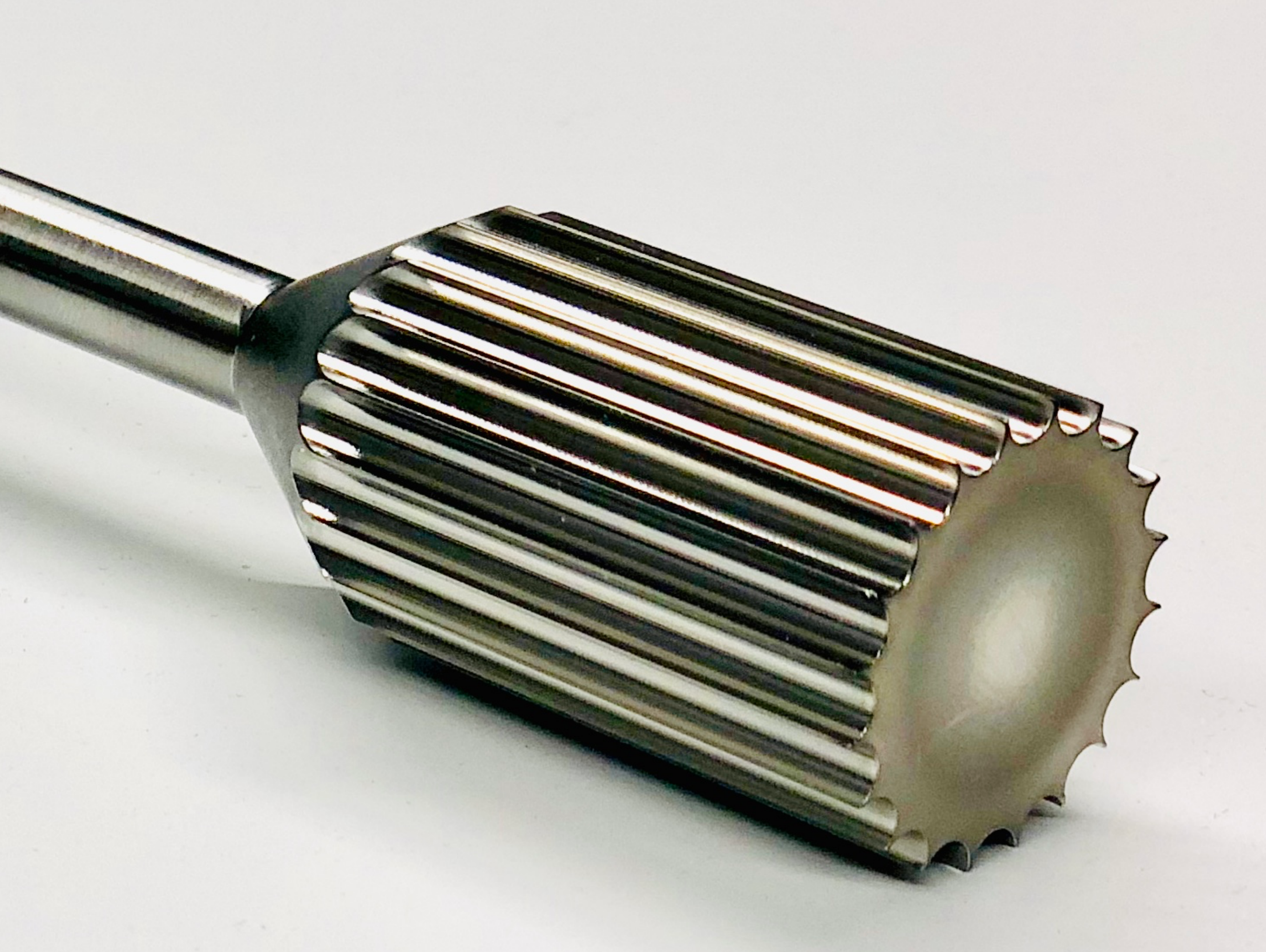}};
\end{scope}

\node[anchor=north west, xshift=4pt, yshift=-5pt] (bbb) at (scope1.north west) {\textbf{(b)}};
\node[anchor=north west,xshift=-122pt] (aaa) at (bbb.north west) {\textbf{(a)}};

\end{tikzpicture}
\caption{\label{fig:rheometerMethod}
Rheometry setup. (a)~Profiled inner cylinder. (b)~Shear cell dimensions (section view), with inner and outer cylinders diameter, $D_{i,\,o}$, respectively. 
}
\end{figure}

\textit{Rheometry.---}The rheometry setup is an open-surface Taylor-Couette (coaxial cylinders, inner cylinder rotating).
The surface of the inner cylinder promotes particle-particle contact during shear (Fig.\,\ref{fig:rheometerMethod}a). 

\textit{Packing fractions.---}
The global packing fraction is determined
as $\varphi(u) = M / \rho_\text{p}V(u)$, where the sample volume
is obtained from 
the fluidization-dependent mean bed height, $h(u)$, determined by image analysis.

The fit function to $\varphi(u)$, shown as inset of Fig.\,\ref{fig:experimental}, 
is of form $\varphi = (A u)^{-B}$, with
$A = \SI{91960}{\s\per\meter}$, $B = 0.068$.

\section{Appendix II:~Strain constants}
In the stationary state, forces in a
fluid must balance, i.e., in terms of the stress tensor,
$\underline{\underline\sigma}$, we have
$\underline\nabla\cdot\underline{\underline\sigma} = 0$. Focusing on
the shear stress $\sigma(r) := \sigma_{r\vartheta}(r)$, neglecting
other shear components and assuming a homogeneous pressure, this reads
\begin{equation}
  \label{eq:5}
  (\underline\nabla\cdot\underline{\underline\sigma})_{\vartheta}
  := \frac{\partial\sigma(r)}{\partial r} + \frac{2\sigma(r)}{r} = 0.
\end{equation}
In terms of the measured stress at the inner cylinder,
$\sigma\equiv\sigma(R_i)$, this implies
$\sigma(r) = (R_i/r)^2\sigma$, i.e., the shear stress decreases
quadratically towards the outer cylinder.

Assuming no-slip boundary conditions, the angular velocity $\varOmega(r)$
is fixed at the cylinders surfaces, $\varOmega(R_i) = \varOmega$ and
$\varOmega(R_o) = 0$. The shear rate is related to the gradient
of the angular velocity, $\dot\gamma(r) = r \,d\varOmega(r)/dr$,
such that
\begin{equation}
  \label{eq:6}
  \varOmega = \int\displaylimits_{R_\mathrm{i}}^{R_\mathrm{o}}dr
  \frac{d\varOmega}{dr}
  = \int\displaylimits_{R_\mathrm{i}}^{R_\mathrm{o}}dr\frac{\dot\gamma(r)}{r}
  = \int\displaylimits_{\sigma/\delta^2}^{\sigma}\frac{\dot\gamma(s)ds}{2s}.
\end{equation}
Using Newtonian rheology, $\dot\gamma(\sigma) = \sigma/\eta$, we find
\begin{equation}
  \label{eq:7}
  \eta = \frac{\delta^2 - 1}{2\delta^2}\times\frac{\sigma}{\varOmega}
\end{equation}
and recover the well-known strain constant for the Taylor-Couette
geometry,
\begin{equation}
  \label{eq:8}
  \dot\gamma_\text{N} = \frac{2\delta^2}{\delta^2 - 1}\varOmega.
\end{equation}
Assuming Bagnold rheology, instead, $\dot\gamma(\sigma) =
\sqrt{\sigma/B}$, Eq.\,(\ref{eq:6}) yields
\begin{equation}
  \label{eq:8}
  B = \frac{(\delta - 1)^2}{\delta^2}\times\frac{\sigma}{\varOmega^2}
\end{equation}
and, respectively, the strain constant
\begin{equation}
  \label{eq:9}
  \dot\gamma_\text{B} = \frac{\delta}{\delta - 1}\varOmega.
\end{equation}

\section{\red{Appendix III: Taylor instability}}
\label{sec:taylor-instability}

For dense granular suspensions sheared in a Taylor-Couette geometry,
the critical Reynolds number above which flow instabilities appear $\Rey_\text{c}\sim\orderof (100)$ \cite{Coussot1999,Majji2018,Kang2021,Ramesh2019,Dash2020,Baroudi2023}.
Throughout our experiment, we strictly find $\Rey < 10$ (see Fig.\ \ref{fig:ReyNumb}),
suggesting circular Couette flow (laminar).

\begin{figure}[h!]
\centering
\input{ReynoldsNumberHeatmap}
\caption{\label{fig:ReyNumb}
Reynolds number, $\Rey$, calculated on the experiment parameter space, \textit{versus} shear rate, $\dot\gamma$, and packing fraction, $\varphi$. Marks are the measured datapoints.}
\end{figure}

The dimensionless Taylor number, $\Ta$,
that compares the Coriolis to viscous forces,
depends, besides the shear geometry,
on the rotation rate, $\varOmega$, and the kinetic
viscosity, $\nu := \eta/\rho_\text{b}\varphi$.
We use the definition proposed by \citeauthor{DiPrima1984} \cite{DiPrima1984}:
 \begin{equation}
   \label{eq:TaylorNbr}
   \Ta = 2 \frac{(\delta-1)}{(\delta+1)} \left(\frac{\varOmega R_i (R_o-R_i)}{\nu} \right)^2 = \kappa \Rey^2.
 \end{equation}
Other definitions have been used \cite{Gillissen2019,Dash2020,Wronski1990,Jastrzkebski1992,Pascal1995};
for $\Ta =\kappa  \Rey^2$, generally $\kappa\sim\orderof (1)$.

The critical Taylor number $\Ta_\mathrm{c}$ depends on 
the nature of the fluid and shear geometry.
The influence of the geometry
is rather well understood in case of Newtonian fluids, 
with wider gaps having a higher $\Ta_\mathrm{c}$ \cite{DiPrima1984}.
For non-Newtonian fluids, $\eta\propto{\dot\gamma}^{R\neq0}$, the dependence
of $\Ta_\mathrm{c}$ on the flow index, $R$, 
is mostly investigated for
shear thinning or at most mildly shear thickening fluids
\cite{Wronski1990,Jastrzkebski1992,Pascal1995}.
We are not aware of
explicit results for $R=1$, relevant for our Bagnold regime,
and particle concentration has yielded contradictory results \cite{Gillissen2019,Dash2020,Baroudi2023}.

For non-Brownian suspensions at relatively high particle concentration ($\varphi=0.5$),
\citeauthor{Dash2020} \cite{Dash2020}
define $\Ta \propto \Rey^2$ and measure $\Ta_\mathrm{c} \sim\orderof (10^5)$.
Others \cite{DiPrima1984,Gillissen2019,Majji2018}
define $\Ta \propto \Rey$
and find that the onset of Taylor instability happens at $\Ta_\text{c} \sim \orderof (50)$, 
translating in our definition to $\Ta_\mathrm{c} \sim\orderof (10^3)$.
For lack of (i)~a 
widely accepted definition of $\Ta$, (ii)~a clear critical value $\Ta_\mathrm{c}$, 
and (iii)~understanding of the evolution of $\Ta_\mathrm{c}$ for non-Newtonian suspensions,
we consider that following our definition, $\Ta < 10^3 \lesssim \Ta_\text{c} $ seem to indicate that the shear thickening regime observed cannot be explained by the Taylor instability.

\section{Appendix IV: Constitutive relation}
\label{sec:constitutive-relation}

The \acrfull{gitt}~\cite{Kranz2018, Kranz2020} formalism takes
  into account the scaling analysis presented in the main text, to
provide a dimensionless constitutive relation for dissipative smooth
hard sphere in terms of a generalized Green-Kubo integral,
\begin{multline}
  \label{eq:constitutiveequation}
  \sigma(\dot\gamma/\omega_0\mid\varphi)/P_0\\
  = \frac{\dot\gamma}{\omega_0} \, \frac{T}{T_0}
  \sum_{\vec q}\int_0^{\infty}d(\omega_0 t)
  \mathcal V_{\vec q\vec q(-t)}\Phi^2_{\vec q(-t)}(t),
\end{multline}
where $P_0$, $\omega_0$ are the ideal particle pressure
and collision frequency in the unsheared fluidized bed, respectively.

The stress-density coupling constant, $V_{\vec q\vec q(-t)}$, is known
explicitly \cite{Kranz2020}. In addition to the explicit shear rate
dependence of the above relation, the shear rate also affects the
advection of the wave vectors, $\vec q(-t)$, and the density
correlator, $\Phi_{\vec q}(t)$, allowing for non-Newtonian rheology
\cite{Fuchs2002,Kranz2020}.

The constitutive equation
(Eq.\ \ref{eq:constitutiveequation}) uses the packing fraction, $\varphi$,
to uniquely characterize the granular fluid. Note however that the
shape of the \gls{gitt} constitutive relation depends weakly on one
more parameter \cite{Kranz2018}, the coefficient of restitution,
$\varepsilon$, which we fix here to $\varepsilon=0.8$.  The granular
temperature, $T$, in the sheared stationary state is determined by the
power density balance,
\begin{equation}
  \label{eq:4}
  \Pi_{\dot\gamma}(T) + \Pi_\mathrm{f} = \Pi_\mathrm{c}(T).
\end{equation}

\end{document}